\documentstyle[aps,twocolumn,prl,epsf]{revtex}

\begin{document}
\draft

\twocolumn[\hsize\textwidth\columnwidth\hsize\csname @twocolumnfalse\endcsname
\title{Weak magnetoresistance of disordered heavy fermion systems}
\author{A.\ Chattopadhyay$^1$, M.\ Jarrell$^1$, H.\ R.\ 
Krishnamurthy$^{1,2}$, H.\ K.\ Ng$^3$, J.\ Sarrao$^4$ and Z.\ Fisk$^5$}
\address{ 
         $^1$ Department of Physics, University of Cincinnati \\
           Cincinnati, Ohio 45221\\
         $^2$ Department of Physics, Indian Institute of Science \\
           Bangalore 560012, India \\
         $^3$ Department of Physics and Center for Materials Research and    
              Technology \\
           Florida State University, Tallahassee, FL 32306-4350 \\
         $^4$ MS K764 / Group MST-10, Los Alamos National Laboratory \\
              Los Alamos, NM 87545 \\
         $^5$ National High Magnetic Field Laboratory \\
             Florida State University, Tallahassee, FL 32306
}
\date{\today}
\maketitle

\widetext
\begin{abstract}
\noindent
We compare the magnetoresistance of UCu$_{3.5}$Pd$_{1.5}$ with 
calculations done within the disordered heavy fermion framework of 
Miranda et al.\ using a phenomenological spectral function for the 
Anderson model, calibrated against Bethe ansatz and quantum Monte Carlo 
results. Both in experiment and theory, we find a weak negative 
magnetoresistance. In contrast, thermodynamic quantities have a 
strong field dependence. Using qualitative arguments broad distribution 
of Kondo scales, we explain the different field dependence of 
susceptibility and resistivity.
\end{abstract}

]

\narrowtext
\paragraph*{Introduction.}
In the past few years, transport and thermodynamic measurements on several
heavy fermion systems have suggested the possibility of a non-fermi
liquid ground state. The violation of 
fermi liquid theory is evident in the measured linear resistivity at low T, 
a logarithmic low temperature divergence of the susceptibility and the 
specific heat coefficient, an optical conductivity with a low frequency 
pseudogap and a linear transport scattering rate at low frequencies
\cite{hfnfl,degiorgi,vonL}. Several 
theoretical models have been proposed to explain this breakdown 
of the fermi liquid paradigm. They include the multichannel Kondo 
impurity and lattice models\cite{cox}, quantum fluctuations near a quantum 
critical point \cite{andraka,tsvelik}, competition between local charge and 
spin fluctuations in infinite dimensions \cite{Si}, and the disordered
Kondo model with a disorder driven 
distributions of Kondo scales \cite{bernal,miranda}.

In this paper we discuss the properties of
the disordered Kondo model in a magnetic field. 
The experimental systems we focus on for comparison are the
the U-based compounds Y$_{1-x}$U$_{x}$Pd$_{3}$ and UCu$_{3.5}$Pd$_{1.5}$  
\cite{hfnfl,andraka} where there is some consensus that the non-Fermi 
liquid ground states are disorder induced. In a magnetic field
both of these systems display 
a very weak negative magnetoresistance, but a large change in 
susceptibility \cite{andraka,bernal}. This is in sharp 
contrast to canonical dilute magnetic alloys with a single small Kondo scale 
(La$_{1-x}$Ce$_{x}$B$_6$ ; $T_{K}\approx1K$) where one usually sees a 
strong negative magnetoresistance ($\Delta\rho(H)/\rho(0)\approx$ 90\% at 
fields $\sim 2-3T$) with a comparable change in the thermodynamic
response. We address these two issues in this paper.
First, we discuss the calculation of magnetoresistance 
within the disordered Kondo model, 
compare the theoretical results with new experimental data 
(reported here for the first time) and
present a qualitative argument for the observed weak field dependence 
of $\rho(H)$ at finite temperatures. Second, we present an explanation for 
the difference in the field dependence of  transport and thermodynamic 
quantities at $T=0$.

\paragraph*{Formalism} The theoretical calculations reported in this paper
use an extension to finite magnetic fields of the formalism
developed in a previous paper \cite{ac}, where the 
optical conductivity for a $\em{dilute}$ collection of Kondo impurities in a 
disordered host, such as in  Y$_{1-x}$U$_{x}$Pd$_{3}$, 
was calculated assuming a distribution of Kondo 
temperatures. However, we believe that the leading dependence (at $T,H$ much 
smaller than the bulk $T_{K}^{0}$) should apply to 
concentrated systems like UCu$_{3.5}$Pd$_{1.5}$ as well.

The dilute system may be modeled in terms of symmetric 
nondegenerate Anderson impurities. In the absence of a magnetic field, the 
impurity f-density of 
states has a sharp resonance at the Fermi level, giving significant 
impurity contributions to the transport, specific heat and magnetic 
susceptibility. 
Most of the heavy fermion systems we consider exhibit strong $L-S$ 
coupling. Hence the dominant effect of the B-field is to 
Zeeman couple to the f-spins. For a given $T_K$, the 
magnetic field splits the up and down f-spins, giving rise to a different 
spectral function for each spin species. 
Concentrating on the Kondo limit of the 
Anderson model ($n_{f}\approx 1$), the phenomenological 
expression \cite{ac,ds,hrk} for the Kondo resonance 
can be extended to include the magnetic 
field effects as 
\begin{equation}
A_{f\uparrow}(\omega)= \frac{1}{\pi\Delta}
{\rm{Re}}
\left[
\frac{i\Gamma_K}{(\omega-\alpha H+i\sqrt{\Gamma_K^2+\beta^2+\gamma^2})}
\right]
^{1/2}
\,.
\end{equation}
Here $\alpha$ and $\beta$ are assumed to be functions of magnetic 
field but independent of temperature,  $\Gamma_{K} = (\pi/2)^{2}T_{K}$,  
$\Delta$ is the f-d hybridization energy and $\gamma(T/T_{K})$ is a 
universal function that captures the temperature dependence of the 
spectrum \cite{ac}. 
The Zeeman-like term $\alpha$ reflects the shift in the spectral peak with 
respect to the Fermi energy. Particle-hole symmetry implies that  
$A_{f,\uparrow}(\omega) = A_{f,\downarrow}(-\omega)$.
  
At $T=0$, the impurity magnetization is given by (in units of 
$\frac{g \mu_{B}}{2}$)
\begin{equation}
m(H/T_{K}) = 
\int_{-\infty}^{0} d\omega 
\left[A_{f\uparrow}(\omega,H/T_{K}) - 
A_{f\downarrow}(\omega,H/T_{K})
\right]
\,.
\end{equation}
When $H/T_{K}<<1$, it is reasonable to assume that to leading order
$\alpha(\frac{H}{T_K})\approx\alpha_0$  and  
$\beta(\frac{H}{T_K})\approx \beta_{0}\frac{H^2}{T_K^2}$ (where the 
prefactors $\alpha_0$ and $\beta_0$ are now constants ). Under these
assumptions, when eq.(2) is expanded up to $O(\frac{H}{T_K})^3$, we have
\begin{equation}
m(H/T_{K}) \approx
\frac{4 T_{K}}{\pi\Delta}
\left[
\frac{\alpha_0}{2}(\frac{H}{T_K}) - 
(\frac{\alpha_0^3}{16}+\frac{\alpha_0\beta_0}{8})(\frac{H}{T_K})^3
\right]
\,.
\end{equation}
By comparing the right hand side of eq.(3) to the Bethe ansatz 
expression for the low-field impurity magnetization in powers of 
$\frac{H}{T_K}$ \cite{andrei} 
we get the low $H$ values of $\alpha, \beta$. 

In principle 
the Bethe ansatz series (which is valid for $H \le T_{K}$) can be 
reproduced only by using an infinite number of coefficients in
our spectral function, but $\alpha_0$ and $\beta_0$ capture the dominant 
low field effects. This is confirmed at a low $T\ne0$ by calculating 
the f-spectra of the symmetric Anderson model in a finite 
field by Quantum Monte Carlo(QMC). The numerical analytic continuation 
is done using maximum entropy method
(MEM) \cite{silver,jarrell_mem}, with the perturbation theory spectrum 
as the high temperature default. Using the MEM results, the 
expression for the spectral function can be calibrated and
$\alpha(H/T_{K})$ and  $\beta(H/T_{K})$ determined
over a wide range of field values. The results are shown in Fig.~1. 
The fits tend to be reasonable until the frequency reaches the charge 
transfer peak.

In the dilute limit, the impurities 
can be assumed to be independent
scatterers and the universal single impurity
spectral function (Eq.1) can be disorder averaged
to calculate dynamical quantities.	
The Hilbert transform of $A_f(\omega,T_{K})$  for each spin species 
then gives the average impurity t-matrix
\begin{equation}
t_f(z)= \int dT_K P(T_K) \int d\omega \frac{V^2 A_f(\omega,T_{K})}{z-\omega}\,,
\end{equation}
where $V$ is the f-d hybridization. For the phenomenological spread of Kondo scales, 
we assume the form 
\begin{equation}
P(T_{K}) = \frac{0.01}{e^{(T_{K}-T_{K}^{0})}+1}\,.
\end{equation} 
This distribution is not based on microscopics; it simply satisfies the experimental criterion of constancy at low $T_{K}$ and looks qualitatively similar to the $P(T_K)$ for UCu$_{5-x}$Pd$_x$\cite{bernal}. Following 
Miranda, et al.\cite{miranda} one uses
the dynamical mean-field approximation\cite{metzvoll}, which becomes 
exact in the limit of infinite dimensions, 
to calculate the lattice self 
energy for a concentration $x$ of substitutional Kondo impurities as
\begin{equation}
\Sigma(\omega)=\frac{x t_f(\omega)}{1+ x t_f(\omega){\cal{G}}(\omega)}\,,
\end{equation}
where ${\cal{G}}(\omega)$ describes the average effective medium
of the impurity.  It is related to the average local greens function
$G$ 
\begin{equation}
G(\omega) = \int d\epsilon \frac{N(\epsilon)}{\omega-\epsilon +\mu -\Sigma(\omega)}\,,
\end{equation}
through the relation 
\begin{equation}
{\cal{G}}^{-1} = G^{-1} + \Sigma\,,
\end{equation}
where $N(\epsilon) = \frac{1}{t^{*}\sqrt\pi} e^{-\epsilon^{2}/t^{*^{2}}}$
and we set $t^{*} = 10,000K$ to establish a unit of energy and temperature. 
The solutions of Eqns.\ 4--8 then give the full self energy.
\par\indent The knowledge of this self energy enables one to calculate
physical quantities\cite{pruschke} like transport coefficients and the 
optical conductivity. In this paper we concentrate on the magnetoresistance 
$\rho(H)$.  
It is measured in units of $\rho_{0}= 2\hbar/e^2\pi a$, 
which (with $h/e^2 \approx 2.6\cdot10^4\Omega$,) varies between 
$10^{2}...10^{3}[(\mu\Omega cm)]$, depending on the lattice 
constant $a$. 

\paragraph*{Results}   

Figs.2a and 3b show some new experimental data for the magnetoresistance
of UCu$_{3.5}$Pd$_{1.5}$ and Figs. 2b and 3a the calculated values.
The experimental measurements were made on a
polycrystalline sample of size 3.0 mm by 0.47 mm by 0.52 mm. 
The sample was grown by arc melting the constituents
and sealed in quartz tube and annealed for 10 days at
800$^o$C. The resistivity was measured using a 
4 probe technique in a 12 Tesla
Oxford superconducting magnet. A constant current of 10 mA was used for 
all the measurements. To eliminate the possibility of any heating effect, 
the current was reduced to 5 mA for one of the temperature scans. No
discernible difference was observed. The calculations were done assuming
$T_{K}^{0} = 100K$ which is roughly the bulk Kondo temperature
in UCu$_{3.5}$Pd$_{1.5}$.

As one can see, there is strong qualitative resemblance between the 
experimental and theoretical curves. The general 
trend is that at very low $T$, there is a negative 
magnetoresistance even at the lowest fields. As one increases $T$, 
$\rho(H)$ is relatively constant up to field values comparable to the 
temperature, beyond which there is a downturn. 
From Fig.~2, we see that there is almost quantitative agreement 
at $4K$. At $20K$, we are at temperatures comparable to $T_{K}^0$ and
the self-consistent modifications to $P(T_{K})$ presumably are 
significant for the lattice. The qualitative agreement is still there 
in that $\rho(H,T=20K)$ is nearly a constant over $12T$.

This interplay of magnetic
field and temperature can be understood as follows. Let us look at the
temperature dependence of the susceptibility and resistivity for a given
Kondo scale. In the case of $\chi(T/T_K)$, the $T=0$ value is given by 
$\frac{1}{T_K}$ , which means that lowering the Kondo temperature will
result in an enhancement of $\chi(T=0)$. A similar
argument holds for $C_{V}/T$ as well. So in a 
system with a distribution of Kondo scales, the low $T$ thermodynamics
($\chi, C_V$) are dominated by the sites with the lowest $T_K$ values.
The case of resistivity is different since for any fixed $T_K$, 
$\rho(T=0)$ is pinned at the unitarity limit. 
Hence lowering of the Kondo scales will end up "sqeezing" the 
"local Fermi liquid" regime in temperature (where $\delta\rho(T/T_{K})$ 
scales like $1-(T/T_{K})^2$). As a result, the dominant
(but by no means exclusive) contribution to transport at low $T$ comes 
from sites with high Kondo scales, particularly $T_{K}\agt T$. 

Now, in the presence of a broad distribution $P(T_{K})$, consider
a temperature $T$ and a field $H<<T$. For any site with 
$T_{K}\alt H$ the Kondo singlet will be broken. Since these low $T_K$ sites 
are the ones dominating the 
thermodynamics, any finite magnetic field will have an appreciable effect 
on $\chi$. On the other hand, a majority of the sites that affect transport 
at $T$ are unaware of this low field. The field dependence of the
resistivity 
shows up only when $H\approx T$, where the system acquires an increasing 
number of free spins by breaking the relevant Kondo singlets. This gives
rise to a weak negative magnetoresistance which is confirmed by the 
magnetotransport measurements on UCu$_{3.5}$Pd$_{1.5}$. The broadness
of $P(T_{K})$ in this case is crucial for $\Delta\rho(H)/\rho(0)$ to
be weak compared to La$_{1-x}$Ce$_{x}$B$_6$, for instance. This is
because even when $H\approx T$, there are a significant number of sites 
that have Kondo scales $T_{K}>>T,H$  that contribute to the resistivity 
through nearly unitary scattering. 

A natural extension of this argument would suggest a gentle initial 
upturn in the resistivity data at $H\approx T$ if there is 
an initial upturn in $P(T_{K})$. This trend is noticed in the data 
on UCu$_{3.5}$Pd$_{1.5}$ [Fig.~2(a),3(b)], indicating that the 
distribution of Kondo scales in the real sample is not really flat
at the low $T_K$ end but has an upward slope. We have calculated 
$\rho(H,T)$ for such a distribution and confirmed this behaviour.  

We have also re-calculated $\chi$ within our framework (earlier 
calculations were done by Bernal 
et al. \cite{bernal} and Miranda et al. \cite{miranda}) to illustrate 
the difference with resistivity in a field and show the results 
in Fig.~3c. Here, for consistency with experiments, $\chi(T)$ is defined 
as $\overline{m(H,T)}/H$ where $\overline{...}$ denotes disorder 
averaging. The 
resistivity responds to the magnetic field much more weakly than $\chi$ 
does. At $T=1K$, $\rho$ changes by $3\%$ when a $6$ Tesla field is applied
to the system; $\chi$ changes by $24\%$ over the same range. This is 
in qualitative agreement with the experiments on Y$_{1-x}$U$_x$Pd$_3$ and 
UCu$_{3.5}$Pd$_{1.5}$. 

The significantly different field dependence of $\rho$ and $\chi$ 
can be understood as follows from the $T=0$ Bethe ansatz expression 
relating the magnetization to the resistivity in the single impurity 
Kondo model:

\begin{equation}
R_{imp}(H/T_K) = R_{imp}(0) \cos^{2} \left[\frac{\pi}{2} f(x) \right].
\end{equation}
Here $x \equiv \frac{g\mu_{B}H}{k_{B}T_{K}} = 
\frac{\tilde{H}}{T_{K}}$ and $f(x) = \frac{2} {g \mu_{B}} m(x)$.
Disorder averaging, for simplicity, over a rectangular distribution of 
$T_K$ upto $T_K^0$, one gets 

\begin{equation}
\overline{\frac{R_{imp}(H)}{R_{imp}(0)}} \sim \frac{\tilde{H}}{T_{K}^{0}} 
\int_{\frac{\tilde{H}}{T_{K}^{0}}}^{\infty}
\cos^{2} \left[\frac{\pi}{2} f(x) \right] \frac{1}{x^2} dx 
\end{equation}

\begin{equation}
\overline{\chi_{imp}(H)} \sim \int_{\frac{\tilde{H}}{T_{K}^{0}}}^{\infty}
f^{\prime}(x) \frac{1}{x} dx \,.
\end{equation}
For large $\tilde{H}$, the integrals are dominated by 
large $x$ where the integrands scale like 
$\frac{\pi^2}{16} \frac{1}{(x \ln(x))^2}$
and  $\frac{1}{2} \frac{1}{(x \ln(x))^2}$ respectively. So the
integrals are comparable for large $\frac{\tilde{H}}{T_{K}^{0}}$, 
implying that $\overline{\chi_{imp}}$ falls off faster than 
$\overline{R_{imp}(H)}$. 
For small $\frac{\tilde{H}}{T_{K}^{0}}$, $\overline{\chi_{imp}}$ has a log
divergence from the small $x$ region. The resistivity integral has a
$\frac{T_{K}^{0}}{\tilde{H}}$ divergence as well, but it gets balanced out 
by the prefactor. Hence, $\overline{R_{imp}(H)}$ has a weaker 
dependence on field compared to $\overline{\chi_{imp}(H)}$ for both large
and small values of $\frac{\tilde{H}}{T_{K}^{0}}$.  

\paragraph*{Conclusion.} We have presented calculations of magnetoresistance
within the disordered Kondo scenario and new experimental data for the 
magnetoresistance in UCu$_{3.5}$Pd$_{1.5}$ at a variety of temperatures  
and find the results to be consistent. The weak magnetoresistance of
this system at finite temperature has been qualitatively explained 
by examining the interplay of field and temperature in the presence of a wide
distribution of Kondo scales. Finally, we have addressed the stronger 
field dependence of susceptibility as compared to transport and presented
a calculation at $T=0$ to explain this.

\par\indent We are thankful to V. Dobrosavljevi\'{c} for motivating
this problem. We would also like to acknowledge useful 
discussions with 
M.\ Hettler,
M.\ Ma,
D.\ E.\ MacLaughlin, 
Anirvan\ Sengupta and 
Q.\ Si.

This work was supported by NSF grants DMR-9406678 and DMR-9357199.

\begin{figure}[t]
\epsfxsize=3.3in
\epsfysize=3.0in
\epsffile{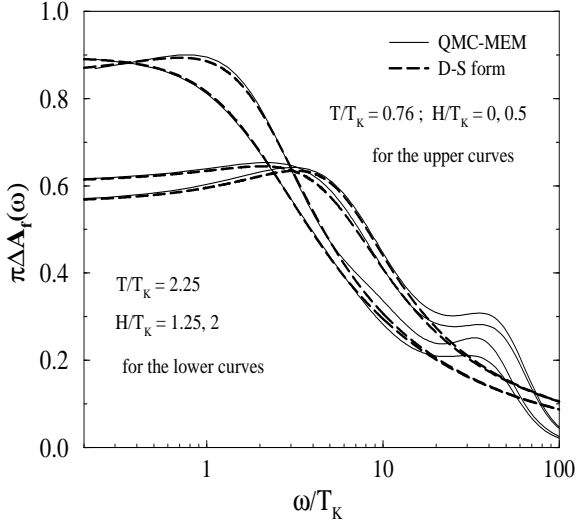}
\caption[]{{Spectral function for the up-spin f-electron in the Kondo 
 limit of the Anderson model for different $T/T_{K}$and $H/T_{K}$ values. 
The dashed lines are the result of the extended Doniach-Sunjic function. 
The solid lines are the QMC/MEM results. The fit is very reasonable until 
the charge transfer physics becomes important. }}     
\end{figure}

\begin{figure}[t]
\epsfxsize=3.8in
\epsfysize=2.25in
\epsffile{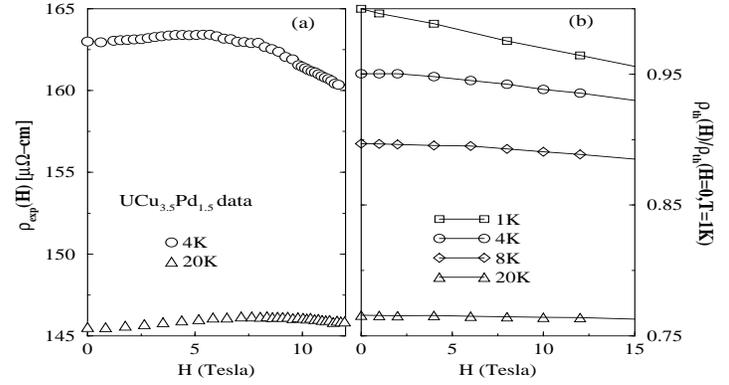}
\caption[]{{Magnetic field dependence of the resistivity. Both the
experimental measurements (a) on UCu$_{3.5}$Pd$_{1.5}$  and theoretical 
plots (b) are shown. The qualitative agreement is quite reasonable. The 
magnetoresistance becomes negative at 
field values comparable to the temperature. This is because transport at 
a given $T$ is dominated 
by sites that have $T_{K}\approx T$ or higher. As a result, at any given
$T$ the resistivity is unaware of the field until $H$ suppresses the 
relevant $T_K$. }}
\end{figure}

\begin{figure}[t]
\epsfxsize=3.8in
\epsfysize=3.0in
\epsffile{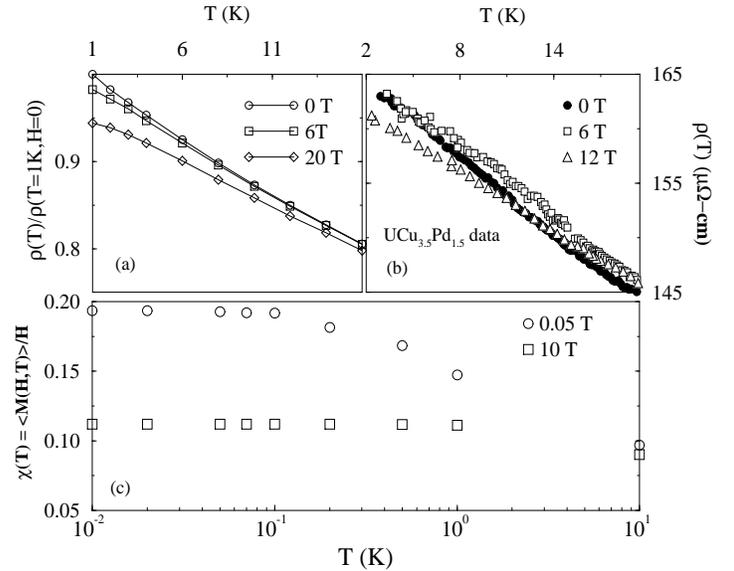}
\caption[]{ Resistivity and impurity spin susceptibility within the 
 disordered Kondo model. The 
 susceptibility shown is normalized with its value at $0.01 K, 
 0.05$ Tesla. One notices a 
 far more pronounced effect of magnetic field on $\chi(T)$ at low 
 temperatures, compared to $\rho(T)$. For instance, at $T = 1K$, $\rho$ is 
 estimated to change by $3 \%$ over 
 $6 T$; $\chi$ changes by $24 \%$ over the same field value. } 
\end{figure}   


\begin{references}
\bibitem{hfnfl} For a recent review, see M.B. Maple {\it et al.}, 
J. Low Temp. Phys. {\bf 99}, 223 (1995), and references cited therein.
\bibitem{degiorgi} L. Degiorgi, H.R. Ott, and F. Hulliger,
Phys.\ Rev.\ B {\bf 52}, 42 (1995); 
L.\ Degiorgi and H.R Ott, J.Phys. :Cond. Matter {\bf 8}, 9901(1996); 
and L.Degiorgi {\it et al.}, Phys.\ Rev.\ B {\bf 54}, 6065 (1996).
\bibitem{vonL} H. v. L\"{o}hneysen {\it et al.}, Phys. Rev. Lett. {\bf 72},
3262 (1994).
\bibitem{cox} a) P. Nozi\`{e}res and A.\ Blandin, J. Phys.
(Paris) {\bf 41},
193 (1980); b) D.\ L.\ Cox, Phys.\ Rev.\ Lett. {\bf 59}, 1240 (1987);
c) Physica B {\bf 186-188}, 312 (1993).
\bibitem{andraka} B.\ Andraka and A.\ M.\ Tsvelik, Phys.\ Rev.\ Lett. 
{\bf 67}, 2886 (1991).
\bibitem{tsvelik} M.\ A.\ Continentino, Phys. Rev. B{\bf 47}, 11587
(1993); A.\ J.\ Millis, Phys. Rev. B{\bf 48}, 7183 (1993); 
A.\ M.\ Tsvelik and M.\ Reizer, Phys. Rev. B{\bf 48}, 9887 (1993).
\bibitem{Si} Q.\ Si, J.Phys. :Cond. Matter {\bf 8}, 9953(1996) and 
references therein.
\bibitem{bernal} O.\ O.\ Bernal {\it et al.}, Phys. Rev. Lett. {\bf 75},
2023 (1995).
\bibitem{miranda} E.\ Miranda, V.\ Dobrosavljevi\'{c} and G.\ Kotliar,
Phys.\ Rev.\ Lett. {\bf 78}, 290 (1997).
\bibitem{ac} A.\ Chattopadhyay and M.\ Jarrell, Phys. Rev. B{\bf 56}, 
R2920 (1997).
\bibitem{ds} S.\ Doniach and M.\ Sunjic, J.Phys.C {\bf 3}, 285 (1970); 
H.\ O.\ Frota and L.\ N.\ Oliveira, Phys.\ Rev.\ B {\bf 33}, 7871 (1986).
\bibitem{hrk} H.\ R.\ Krishna-murthy, J.\ W.\ Wilkins and 
K.\ G.\ Wilson,  
Phys.\ Rev.\ B {\bf 21}, 1003 (1980); {\bf 21}, 1044 (1980).
\bibitem{andrei} N.\ Andrei {\it et al.}, Rev. Mod. Phys. {\bf 55},
331 (1983); A.\ M.\ Tsvelik and P.\ B.\ Wiegmann, Adv. Phys. {\bf 32}, 
453 (1983).
\bibitem{silver} R.N. Silver {\it et al.}, Phys. Rev. Lett. {\bf 65},
496 (1990).
\bibitem{jarrell_mem} For a review of the Maximum Entropy method of
analytic continuation, see M.\ Jarrell, J.E.\ Gubernatis,  Physics 
Reports Vol.\ {\bf{269}} \#3, (May, 1996).
\bibitem{metzvoll} W. Metzner and D Vollhardt, Phys. Rev. Lett.
{\bf 62}, 324 (1989).
\bibitem{pruschke} Th. Pruschke, M. Jarrell and J.K. Freericks, 
Advances in Physics {\bf{44}}, 187-210 (March/April 1995).
\end{references}
\end{document}